\documentclass[12pt]{iopart}

\usepackage{iopams}
\usepackage{dsfont}
\usepackage{graphics}

\newcommand{\ket}[1]{\left|{#1}\right\rangle}
\newcommand{\bra}[1]{\left\langle{#1}\right|}
\newcommand{\braket}[2]{\left\langle{#1}|{#2}\right\rangle}

\newcommand{\Dcirc}{\mathcal{D}^{\circ}}

\begin{document}

\title{Evolution equation for multi-photon states in turbulence}

\author{Filippus S. Roux}
\ead{froux@nmisa.org}
\address{National Metrology Institute of South Africa, Meiring Naud{\'e} Road, Brummeria 0040, Pretoria, South Africa}
\address{School of Physics, University of the Witwatersrand, Johannesburg 2000, South Africa}

\begin{abstract}
The recently developed Wigner functional theory is used to formulate an evolution equation for arbitrary multi-photon states, propagating through a turbulent atmosphere under arbitrary conditions. The resulting evolution equation, which is obtained from an infinitesimal propagation approach, is in the form of a Fokker-Planck equation for the Wigner functional of the state and therefore incorporates functional derivatives. We show consistency with previously obtained solutions from different approaches and consider possible ways to find additional solutions for this equation.

\end{abstract}



\section{\label{intro}Introduction}

The effect of a turbulent atmosphere on a photonic quantum state during propagation plays an important role in any free-space quantum information system (FS-QIS), such as free-space quantum communication \cite{gisin,krenn2}, quantum teleportation over free-space \cite{telezeil,fstelez,advtelport}, and quantum protocols \cite{homsync2} for free-space time-transfer \cite{sync1}. The random phase modulations induced by the fluctuating refractive index distort the spatial characteristics of the state, which in turn affects the photon-number characteristics of the state.

Current implementations of such FS-QISs focus on either the spatiotemporal degrees of freedom of a few discrete photons \cite{qturb4,turboam2,tyler,qturb3,malik,oamturb,leonhard,qkdturb,qutrit}, or the particle-number degrees of freedom (so-called continuous variables) for a few discrete spatial modes \cite{milburn,contvar1,multiphoton,wasil,sanders,weedbrook,contvar2}. The implementations that are based on the spatial degrees of freedom often use orbital angular momentum (OAM) modes to encode the information \cite{vallone,allen,mtt,cryptotwist}. The polarization degree of freedom has been incorporated in both trends \cite{vallone,kwiat,heim}.

The natural culmination of these endeavors would be an approach that incorporates both the spatiotemporal degrees of freedom and the particle-number degrees of freedom to their full extent. For such an implementation, it would be necessary to understand the evolution during propagation through a turbulent atmosphere of a photonic state that is arbitrary in terms of both its spatiotemporal degrees of freedom and particle-number degrees of freedom.

Such an analysis needs a special formalism in which neither the spatiotemporal degrees of freedom, nor the particle-number degrees of freedom, are treated as discrete variables --- a few discrete photons or a few discrete modes. While the multi-photon nature of the states is represented by the particle-number degrees of freedom, the turbulent medium affects the spatial degrees of freedom of the state, scattering any input mode into an infinite dimensional modal space. This modal scattering process happens continuously, resulting in optical power being scattered back and forth among an infinite number of modes. A proper analysis of the evolution of multi-photon states, propagating through a turbulent atmosphere, therefore requires a formalism that incorporates both the particle-number degrees of freedom and the spatiotemporal degrees of freedom as continuous degrees of freedom.

Such a special formalism has recently been developed in the form of a Wigner functional formalism \cite{wigfunk}. It is a generalization of the well-known Wigner function \cite{wigner}, which has been used extensively (see \cite{wigreview} and references therein). The Wigner functional formalism is based on the spatiotemporal quadrature bases that are eigenstates of the momentum dependent quadrature operators \cite{stquad}. This formalism is an extension of the Moyal formalism for quantum mechanics \cite{groenewold,moyal,psqm}. The Wigner functionals are defined in a phase space that represents a Hilbert space for all possible states in quantum optics. The operations are represented by a functional version of the star-product of the Moyal formalism.

The effect of turbulence on photonic quantum states has been studied extensively, but, as with the implementations, these studies invariably only focused on one of the two sets of continuous degrees of freedom. Those studies that focus on the spatial degrees of freedom \cite{sr,qturb4,turboam2,qturb3,pors,malik,leonhard}, are mostly based on the single-phase screen approximation \cite{paterson}, which assumes weak scintillation \cite{turbsim}. However, we have also considered a multi-phase screen approach (based on infinitesimal propagation) to derive an evolution equation for the state \cite{ipe,iperr,lindb,notrunc,kormed}, which can address strong scintillation conditions \cite{numipe}. Since the scintillation process directly affects the spatial degrees of freedom, the studies that focus on the particle-number degrees of freedom need a method to represent the effect of turbulence on the spatial degrees of freedom. They get around this problem by modeling the effect of turbulence simply as a loss mechanism, ignoring any other role that the spatial degrees of freedom may play in the process \cite{semenov,vasyl}. Another approach, which employs a Wigner function for the spatiotemporal degrees of freedom, instead of the particle-number degrees of freedom, represents the effect of turbulence as a random position dependent force \cite{wig1,wig2,wig3}.

In this paper, we'll use the Wigner functional formalism for an infinitesimal propagation approach to derive an evolution equation for the Wigner functional of an arbitrary photonic state. The result is a functional differential equation in the form of a Fokker-Planck equation, which we call the infinitesimal propagation functional equation (IPFE). This is the main result of the paper.

The IPFE is difficult to solve. We discuss a few approaches for finding solutions. One of these approaches only provide trivial solutions, the vacuum and thermal states. We show that previous results obtained from considering states with a few discrete photons \cite{notrunc,kormed} are also solutions of the IPFE. An interesting observation emerges from considering the characteristic functional of the solution, namely that it obeys an identical functional differential equation as the Wigner functional solution.

\section{\label{ipestat}Infinitesimal propagation}

\subsection{Unitary evolution}

Under general scintillation conditions, a photonic quantum state encounters two effects while propagating through the atmosphere. One is the free-space propagation process (without the effect of turbulence). The other is a continuous phase modulation caused by fluctuations in the refractive index of the medium. While the latter involves a modulation in the position domain, the former represents a modulation in the Fourier domain. The full dynamics, which incorporates both these effects, is captured in the classical context, by the equation of motion for paraxial propagation through a turbulent medium
\begin{equation}
\nabla_T^2 g(\mathbf{x},z) - \rmi 2k\partial_z g(\mathbf{x},z) + 2k^2 \tilde{n}(\mathbf{x},z) g(\mathbf{x},z) = 0 .
\label{eomturb}
\end{equation}
Here $g(\mathbf{x},z)$ is the classical scalar electromagnetic field, $k=2\pi/\lambda$ is the wavenumber and $\tilde{n}(\mathbf{x},z)$ is the fluctuation in the refractive index of the atmosphere. The refractive index of the atmosphere can thus be represented as $n=1+\tilde{n}$. In the transverse Fourier domain, the equation becomes
\begin{equation}
\partial_z G(\mathbf{a},z) = \frac{\rmi 2\pi^2}{k}|\mathbf{a}|^2 G(\mathbf{a},z) -\rmi k \int N(\mathbf{a}-\mathbf{a}',z) G(\mathbf{a}',z)\ \rmd ^2a' ,
\label{fteomturb}
\end{equation}
where $\mathbf{a}$ is the two-dimensional transverse spatial frequency vector, $G(\mathbf{a},z)$ is the transverse angular spectrum, which is also a function of the propagation distance $z$, and $N(\mathbf{a},z)$ is the transverse Fourier transform of the refractive index fluctuation. Moreover, $N^*(\mathbf{a},z)=N(-\mathbf{a},z)$, because the refractive index fluctuation is a real-valued function. Unless the field represented by the photonic state is strong enough to induce a nonlinear interaction with the atmosphere, the classical description of the propagation process, as given in (\ref{eomturb}), suffices to provide the basis for the development of the evolution of photonic states propagating through the atmosphere.

The approach that we'll follow is to derive a dynamical equation from an infinitesimal propagation of the state, as proposed in \cite{ipe,lindb}. If $\hat{U}(z)$ represents that unitary evolution operator for the propagation of a state over a distance $z$ through the atmosphere and $\hat{\rho}$ is the density operator for the input state, then the output state is given by $\hat{\rho}(z) = \hat{U}(z)\hat{\rho}\hat{U}^{\dag}(z)$. For an infinitesimal propagation, one expects to find that $\hat{U}(dz) = \mathbf{1}+\rmi dz \hat{P}_{\Delta}$, where $\hat{P}_{\Delta}$ represents the infinitesimal propagation. It thus leads to an evolution equation of the form $\partial_z \hat{\rho}(z) = \rmi [\hat{P}_{\Delta},\hat{\rho}]$. The infinitesimal propagation operator $\hat{P}_{\Delta}$ assumes full knowledge of the medium that causes the scintillation process. Since we normally only know the statistical properties of the medium, we'll eventually have to apply an ensemble averaging. However, we start by deriving an expression for the infinitesimal propagation operator, using the classical equation in (\ref{fteomturb}).

To define quantum operators for the scintillation process, it is convenient to combine the two terms in (\ref{fteomturb}), so that
\begin{equation}
\partial_z G(\mathbf{a},z) = \rmi \int M(\mathbf{a},\mathbf{a}',z) G(\mathbf{a}',z)\ \rmd ^2a' = \rmi M(z)\diamond G(z) ,
\label{fteomturbs}
\end{equation}
where the kernel is given by
\begin{equation}
M(\mathbf{a},\mathbf{a}',z) \equiv \pi\lambda |\mathbf{a}|^2 \delta(\mathbf{a}-\mathbf{a}') - k N(\mathbf{a}-\mathbf{a}',z) ,
\label{defms}
\end{equation}
and where we adopt the notation introduced in \cite{wigfunk} to write the integral contraction as a binary operation denoted by $\diamond$. Note that the kernel is Hermitian $M^*(\mathbf{a},\mathbf{a}',z) = M(\mathbf{a}',\mathbf{a},z)$. Represented as an infinitesimal propagation, the equation in (\ref{fteomturbs}) becomes
\begin{equation}
G(\mathbf{a},z+dz) = \int \left[\delta(\mathbf{a}-\mathbf{a}') + \rmi dz M(\mathbf{a},\mathbf{a}',z) \right] G(\mathbf{a}',z) \ \rmd ^2a' .
\label{infprops}
\end{equation}

The transverse angular spectrum is interpreted as a Fourier domain wave function $G(\mathbf{a},z) = \braket{\mathbf{a}}{G(z)}$, where $\ket{\mathbf{a}}$ is an element of the transverse spatial frequency basis, which obeys an orthogonality condition $\braket{\mathbf{a}}{\mathbf{a}'}=\delta(\mathbf{a}-\mathbf{a}')$. It allows us to define the single-photon states as
\begin{equation}
\ket{G(z)} = \int \ket{\mathbf{a}}G(\mathbf{a},z)\ \rmd ^2a .
\end{equation}
The infinitesimal propagation for a single-photon state is $\ket{G(z+dz)}=\hat{\mathcal{P}}_1(z,dz)\ket{G(z)}$, where $\hat{\mathcal{P}}_1(z,dz)\equiv\mathbf{1}+\rmi dz \hat{M}(z)$, with
\begin{equation}
\mathbf{1} \equiv \int \ket{\mathbf{a}}\bra{\mathbf{a}}\ \rmd ^2a ,
\end{equation}
being the projection operator for single-photon states, and
\begin{equation}
\hat{M}(z) \equiv \int \ket{\mathbf{a}}M(\mathbf{a},\mathbf{a}',z)\bra{\mathbf{a}'}\ \rmd ^2a\ \rmd ^2a' ,
\end{equation}
representing the infinitesimal propagation process for single-photon states, which also incorporates a projection operation for single-photon states. Since $\hat{M}(z)$ is Hermitian, it follows that $\hat{\mathcal{P}}_1^{\dag}(z,dz)=\mathbf{1}-\rmi dz \hat{M}(z)$.

While $\hat{\mathcal{P}}_1$ implies a projection operation for single-photon states, the equivalent for $n$ photons, expressed as $\hat{\mathcal{P}}_n\equiv\hat{\mathcal{P}}_1^{\otimes n}$, would involve a projection operation for $n$-photon states. For general multi-photon states, one needs to sum over all such operators. Therefore, the unitary operator is given by
\begin{eqnarray}
\hat{U}(z,dz) &= \sum_n \frac{1}{n!} \left[ \hat{\mathcal{P}}_1(z,dz)\right]^{\otimes n} = \exp_{\otimes}[\hat{\mathcal{P}}_1(z,dz)] \nonumber \\
 &= \mathds{1}+\rmi dz \hat{M}(z)\otimes\mathds{1} + \Or \left\{dz^2\right\} ,
\label{unievol}
\end{eqnarray}
where
\begin{equation}
\mathds{1} = \sum_n \frac{\mathbf{1}^{\otimes n}}{n!} = \exp_{\otimes}(\mathbf{1})
\end{equation}
is the identity operator for the complete space, and $\exp_{\otimes}(\cdot)$ implies that all products in its expansion are tensor products and the first term in the expansion contains only the vacuum state $\ket{\rm vac}\bra{\rm vac}$. The last line in (\ref{unievol}) follows under infinitesimal propagation conditions due to the smallness of $dz$. The sub-leading order term serves as the {\em scintillation operator}, $\hat{\mathcal{M}}(z) \equiv \hat{M}(z)\otimes\mathds{1}$. Although it does not involve any projection operations, only one photon is affected by the scintillation process.

For an arbitrary {\em pure} state, the infinitesimal propagation gives
\begin{equation}
\ket{\psi(z+dz)} = \hat{U}(z,dz) \ket{\psi(z)} = \ket{\psi(z)} + \rmi dz \hat{\mathcal{M}}(z)\ket{\psi(z)} .
\end{equation}
In the limit $dz\rightarrow 0$, it leads to the dynamical equations
\begin{equation}
\eqalign{
\partial_z \ket{\psi(z)} = \rmi \hat{\mathcal{M}}(z)\ket{\psi(z)} , \\
\partial_z \bra{\psi(z)} = -\rmi \bra{\psi(z)} \hat{\mathcal{M}}(z) .}
\end{equation}
An arbitrary (pure or mixed) state, expressed as a density operator $\hat{\rho}(z)$, would thus obey the dynamical equation
\begin{equation}
\partial_z \hat{\rho}(z) = \rmi \left[\hat{\mathcal{M}}(z),\hat{\rho}(z)\right] .
\label{unidyns}
\end{equation}
We can now identify the scintillation operator $\hat{\mathcal{M}}(z)$ with the infinitesimal propagation operator $\hat{P}_{\Delta}$, mentioned above. As such, it describes the unitary evolution of the state propagating through the turbulent atmosphere. However, the detail of the medium is only known in a statistical sense. To have a predictive ability, one needs to consider the ensemble average of the process. For that purpose, it is more convenient to consider the state in terms of its Wigner functional.

\subsection{Wigner functional approach}

To make the expression in (\ref{unidyns}) more analytically manageable, we define an auxiliary function for the scintillation operator
\begin{equation}
\hat{\mathcal{R}}(z;\eta) = \exp_{\otimes}\left[ \int \ket{\mathbf{a}} T(\mathbf{a},\mathbf{a}',z;\eta) \bra{\mathbf{a}'}\ \rmd ^2a\ \rmd ^2a' \right] ,
\label{rgendef}
\end{equation}
where $\eta$ is an auxiliary variable, and
\begin{equation}
T(\mathbf{a},\mathbf{a}',z;\eta) \equiv \exp\left(\rmi\pi\eta\lambda|\mathbf{a}|^2\right) \delta(\mathbf{a}-\mathbf{a}') -\rmi\eta k N(\mathbf{a}-\mathbf{a}',z) .
\label{tkerndef}
\end{equation}
The auxiliary variable $\eta$ carries the dimension of a distance and it effectively takes over the role of $dz$. The scintillation operator is recovered by
\begin{equation}
\partial_{\eta} \left. \hat{\mathcal{R}}(z;\eta)\right|_{\eta=0} = \rmi \hat{\mathcal{M}}(z) .
\end{equation}
The dynamical equation can then be expressed as
\begin{equation}
\partial_z \hat{\rho}(z) = \partial_{\eta} \left[\hat{\mathcal{R}}(z;\eta)\hat{\rho}(z)\hat{\mathcal{R}}^{\dag}(z;\eta)\right]_{\eta=0} .
\label{unidyngs}
\end{equation}

The Wigner functional representation of the dynamical equation in (\ref{unidyns}) or (\ref{unidyngs}) requires a Wigner functional for the auxiliary function of the scintillation operator $\hat{\mathcal{R}}(z;\eta)$, which represents a linear operation of the form given in (\ref{rgendef}) and (\ref{tkerndef}). The Wigner functional for such a linear process is given by
\begin{equation}
W_{\hat{\mathcal{R}}}[\alpha] = \frac{1}{\det \{\mathbf{1}+T\}} \exp\left[- 2\alpha^*\diamond (\mathbf{1}-T)\diamond(\mathbf{1}+T)^{-1}\diamond\alpha \right] ,
\label{wiglin}
\end{equation}
under the assumption that $\mathbf{1}+T = \delta(\mathbf{a}-\mathbf{a}')+T(\mathbf{a},\mathbf{a}',z)$ is invertable. The derivation of (\ref{wiglin}) is provided in \ref{linop}.

Note that $(\mathbf{1}+T)|_{\eta=0} = 2\delta(\mathbf{a}-\mathbf{a}')$ and $(\mathbf{1}-T)|_{\eta=0}=0$. It implies that $(\mathbf{1}+T)^{-1}|_{\eta=0} = \delta(\mathbf{a}-\mathbf{a}')/2$. Therefore, if we define $B \equiv 2 (\mathbf{1}-T)\diamond(\mathbf{1}+T)^{-1}$, then, by applying derivatives with respect to $\eta$ and setting $\eta=0$, we get $\partial_{\eta} B|_{\eta=0} = -\rmi M(\mathbf{a},\mathbf{a}',z)$ and $\partial_{\eta} B^*|_{\eta=0} = \rmi M(\mathbf{a}',\mathbf{a},z)$, because $M$ is Hermitian.

In terms of Wigner functionals, the dynamical equation in (\ref{unidyngs}) is represented by\footnote{The Wigner functional for the product of operators is obtained by expressing the operators via the Weyl transformation in terms of their respective Wigner functionals. The result leads to the functional version of the star-product of the Moyal formalism \cite{moyal}.}
\begin{eqnarray}
\fl \partial_z W_{\hat{\rho}}(z)[\alpha] &=& \left. \partial_{\eta} W_{\hat{\mathcal{R}}\hat{\rho}\hat{\mathcal{R}}^{\dag}}[\alpha] \right|_{\eta=0} \nonumber \\
&=& \partial_{\eta} \int W_{\hat{\mathcal{R}}}\left[\frac{\alpha_a}{2}+\frac{\alpha_b}{2}+\frac{\alpha}{2}\right] W_{\hat{\rho}}(z)[\alpha_a]
W_{\hat{\mathcal{R}}^{\dag}}\left[\frac{\alpha_a}{2}-\frac{\alpha_b}{2}+\frac{\alpha}{2}\right] \nonumber \\
&& \left. \times \exp[(\alpha^*-\alpha_a^*)\diamond\alpha_b-\alpha_b^*\diamond(\alpha-\alpha_a)]\ \mathcal{D}[\alpha_a,\alpha_b] \right|_{\eta=0} \nonumber \\
&=& \partial_{\eta} \int \exp\left[ -\frac{1}{4}\left(\alpha_a^*+\alpha^*+\alpha_b^*\right)\diamond B \diamond\left(\alpha_a+\alpha+\alpha_b\right)\right] \nonumber \\
 && \times W_{\hat{\rho}}(z)[\alpha_a] \exp\left[ -\frac{1}{4}\left(\alpha_a+\alpha-\alpha_b\right)\diamond B^* \diamond\left(\alpha_a^*+\alpha^*-\alpha_b^*\right)\right] \nonumber \\
 && \left. \times\exp\left[\left(\alpha^*-\alpha_a^*\right)\diamond\alpha_b-\alpha_b^*\diamond\left(\alpha-\alpha_a\right)\right]\ \mathcal{D}[\alpha_a,\alpha_b] \right|_{\eta=0} ,
\end{eqnarray}
where $W_{\hat{\rho}}(z)[\alpha]$ indicates that the Wigner functional is a {\em function} of $z$ and a {\em functional} of $\alpha$. Evaluating the derivative with respect to $\eta$ and setting $\eta=0$, we get
\begin{eqnarray}
\fl \partial_z W_{\hat{\rho}}(z)[\alpha]
&=& \int W_{\hat{\rho}}(z)[\alpha_a] \left[\frac{\rmi}{2}\left(\alpha_a^*+\alpha^*\right)\diamond M\diamond\alpha_b
 +\frac{\rmi}{2}\alpha_b^*\diamond M\diamond\left(\alpha_a+\alpha\right)\right] \nonumber \\
&& \times \exp\left[\alpha_b^*\diamond\left(\alpha_a-\alpha\right)-\left(\alpha_a^*-\alpha^*\right)\diamond\alpha_b\right]\ \mathcal{D}[\alpha_a,\alpha_b] .
\end{eqnarray}
Again, we use an auxiliary function to express the integrand in terms of an exponential function so that we can evaluate the functional integrals
\begin{eqnarray}
\fl \partial_z W_{\hat{\rho}}(z)[\alpha]
=& \partial_{\xi} \int W_{\hat{\rho}}(z)[\alpha_a] \exp \left[ \alpha^*\diamond\left(\mathbf{1}+\frac{\rmi\xi}{2} M\right)\diamond\alpha_b
-\alpha_a^*\diamond\left(\mathbf{1}-\frac{\rmi\xi}{2} M\right)\diamond\alpha_b \right. \nonumber \\
& \left. \left. \alpha_b^*\diamond\left(\mathbf{1}+\frac{\rmi\xi}{2} M\right)\diamond\alpha_a
-\alpha_b^*\diamond\left(\mathbf{1}-\frac{\rmi\xi}{2} M\right)\diamond\alpha \right]\ \mathcal{D}[\alpha_a,\alpha_b] \right|_{\xi=0} ,
\label{tussen0}
\end{eqnarray}
where $\xi$ is the new auxiliary variable. The argument of the exponential function is linear in $\alpha_b$ and $\alpha_b^*$. Therefore, the functional integration over $\alpha_b$ converts the exponential function into a Dirac delta functional, which replaces $\alpha_a$ in the argument of the state in terms of $\alpha$ and $M$.

To perform the functional integration over $\alpha_b$, we first perform a change of integration variables
\begin{equation}
\eqalign{
\alpha_b \rightarrow \left(\mathbf{1}-\frac{\rmi\xi}{2} M\right)^{-1}\diamond\alpha_c , \\
\alpha_b^* \rightarrow \alpha_c^*\diamond\left(\mathbf{1}+\frac{\rmi\xi}{2} M\right)^{-1} . }
\end{equation}
The functional integrations over $\alpha_b$ and $\alpha_a$ then leads to
\begin{equation}
\partial_z W_{\hat{\rho}}(z)[\alpha]
 = \left. \partial_{\xi} \frac{W_{\hat{\rho}}(z)\left[ \left(2+\rmi\xi M\right)^{-1}\diamond\left(2-\rmi\xi M\right) \diamond\alpha\right]}{\sqrt{\det \left\{1+\frac{1}{4}\xi^2 M^2\right\}}} \right|_{\xi=0} .
\end{equation}

Evaluating the derivative with respect to $\xi$ and setting $\xi=0$, we note that
\begin{equation}
\left. \partial_{\xi} \left( \det \left\{1+\frac{1}{4}\xi^2 M^2\right\}\right)^{-1/2} \right|_{\xi=0} = 0 .
\end{equation}
Hence, the first-order unitary dynamics for the multi-photon state propagating through a turbulent atmosphere, is obtained
\begin{equation}
\partial_z W_{\hat{\rho}} = \rmi\alpha^*\diamond M\diamond\frac{\delta
 W_{\hat{\rho}}}{\delta\alpha^*}-\rmi\frac{\delta W_{\hat{\rho}}}{\delta\alpha}\diamond M\diamond\alpha ,
\label{uniwigs}
\end{equation}
where we dropped the dependences on $z$ and $\alpha$.

The equation in (\ref{uniwigs}) would not provide the required dynamics when we apply ensemble averaging. The reason is that the refractive index fluctuation has a zero mean, which implies that
\begin{equation}
\langle M \rangle =\pi\lambda|\mathbf{a}|^2\delta(\mathbf{a}-\mathbf{a}') .
\label{eam0}
\end{equation}
As a result, an ensemble averaging applied to (\ref{uniwigs}), gives
\begin{equation}
\partial_z W_{\hat{\rho}} = \rmi\pi\lambda\alpha^*\diamond|\mathbf{a}|^2\mathbf{1}\diamond\frac{\delta W_{\hat{\rho}}}{\delta\alpha^*}
-\rmi\pi\lambda\frac{\delta W_{\hat{\rho}}}{\delta\alpha}\diamond|\mathbf{a}|^2\mathbf{1}\diamond\alpha ,
\label{fsdv}
\end{equation}
where $|\mathbf{a}|^2\mathbf{1} \equiv |\mathbf{a}|^2\delta(\mathbf{a}-\mathbf{a}')$. The resulting equation in (\ref{fsdv}) represents free-space propagation without turbulence.

\subsection{Second order}

To see the effect of the turbulence after an ensemble averaging, one must go to second order in the scintillation operator. For this purpose, (\ref{uniwigs}) is first integrated over $z$:
\begin{equation}
\fl W_{\hat{\rho}}(z) = W_{\hat{\rho}}(z_0) + \rmi \int_{z_0}^z \alpha^*\diamond M(z_1)\diamond\frac{\delta W_{\hat{\rho}}(z_1)}{\delta\alpha^*}
- \frac{\delta W_{\hat{\rho}}(z_1)}{\delta\alpha}\diamond M(z_1)\diamond\alpha\ \rmd z_1 .
\end{equation}
Then we substitute the equation back into itself (making the Fourier domain integrals explicit) to obtain the second-order equation
\begin{eqnarray}
\fl W_{\hat{\rho}}(z) =& W_{\hat{\rho}}(z_0) - \rmi \int_{z_0}^z \int \frac{\delta W_{\hat{\rho}}(z_0)}{\delta\alpha(\mathbf{a}_1)} M(\mathbf{a}_1,\mathbf{a}_2,z_1)\alpha(\mathbf{a}_2)\ \rmd ^2 a_1\ \rmd ^2 a_2\ \rmd z_1 \nonumber \\
& + \rmi \int_{z_0}^z \int \alpha^*(\mathbf{a}_1) M(\mathbf{a}_1,\mathbf{a}_2,z_1) \frac{\delta W_{\hat{\rho}}(z_0)}{\delta\alpha^*(\mathbf{a}_2)}\ \rmd ^2 a_1\ \rmd ^2 a_2\ \rmd z_1 \nonumber \\
& - \int_{z_0}^z \int_{z_0}^{z_1} \int \left[ \frac{\delta W_{\hat{\rho}}(z_0)}{\delta\alpha(\mathbf{a}_3)} M(\mathbf{a}_3,\mathbf{a}_4,z_2)
 \delta(\mathbf{a}_4-\mathbf{a}_1) M(\mathbf{a}_1,\mathbf{a}_2,z_1)\alpha(\mathbf{a}_2) \right. \nonumber \\
& + \alpha^*(\mathbf{a}_1) M(\mathbf{a}_1,\mathbf{a}_2,z_1) \delta(\mathbf{a}_3-\mathbf{a}_2) M(\mathbf{a}_3,\mathbf{a}_4,z_2)
\frac{\delta W_{\hat{\rho}}(z_0)}{\delta\alpha^*(\mathbf{a}_4)} \nonumber \\
& + \frac{\delta^2 W_{\hat{\rho}}(z_0)}{\delta\alpha(\mathbf{a}_1)\delta\alpha(\mathbf{a}_3)} M(\mathbf{a}_3,\mathbf{a}_4,z_2)\alpha(\mathbf{a}_4) M(\mathbf{a}_1,\mathbf{a}_2,z_1)\alpha(\mathbf{a}_2) \nonumber \\
& - \alpha^*(\mathbf{a}_3) M(\mathbf{a}_3,\mathbf{a}_4,z_2) \frac{\delta^2 W_{\hat{\rho}}(z_0)}{\delta\alpha(\mathbf{a}_1)\delta\alpha^*(\mathbf{a}_4)} M(\mathbf{a}_1,\mathbf{a}_2,z_1)\alpha(\mathbf{a}_2) \nonumber \\
& - \alpha^*(\mathbf{a}_1) M(\mathbf{a}_1,\mathbf{a}_2,z_1) \frac{\delta^2 W_{\hat{\rho}}(z_0)}{\delta\alpha^*(\mathbf{a}_2)\delta\alpha(\mathbf{a}_3)} M(\mathbf{a}_3,\mathbf{a}_4,z_2)\alpha(\mathbf{a}_4) \nonumber \\
& \left. + \alpha^*(\mathbf{a}_1) M(\mathbf{a}_1,\mathbf{a}_2,z_1) \alpha^*(\mathbf{a}_3) M(\mathbf{a}_3,\mathbf{a}_4,z_2) \frac{\delta^2 W_{\hat{\rho}}(z_0)}{\delta\alpha^*(\mathbf{a}_2)\delta\alpha^*(\mathbf{a}_4)} \right] \nonumber \\
& \times \rmd ^2 a_1\ \rmd ^2 a_2\ \rmd ^2 a_3\ \rmd ^2 a_4\ \rmd z_2\ \rmd z_1 ,
\label{tweede1}
\end{eqnarray}
In effect, we performed the back-substitution twice, but only kept up to second-order terms, because the third-order terms are at least second order in $dz$. As a result, the $z$-dependences of all Wigner functionals are given as $z_0$.

\subsection{Ensemble averaging}

One can now perform the ensemble averaging process. The zeroth-order terms are unaffected. The first-order terms are given by (\ref{eam0}). For the second-order terms, the ensemble averaging involves
\begin{eqnarray}
& \int_{z_0}^z \int_{z_0}^{z_1} \langle M(\mathbf{a}_1,\mathbf{a}_2,z_1) M(\mathbf{a}_3,\mathbf{a}_4,z_2)\rangle\ \rmd z_2\ \rmd z_1 \nonumber \\
=& \int_{z_0}^z \int_{z_0}^{z_1} \pi^2\lambda^2 |\mathbf{a}_1|^2 |\mathbf{a}_3|^2 \delta(\mathbf{a}_1-\mathbf{a}_2) \delta(\mathbf{a}_3-\mathbf{a}_4) \nonumber \\
 & + k^2 \langle N(\mathbf{a}_1-\mathbf{a}_2,z_1) N(\mathbf{a}_3-\mathbf{a}_4,z_2)\rangle\ \rmd z_2\ \rmd z_1 ,
\label{eamm}
\end{eqnarray}
where we substituted in (\ref{defms}) and removed terms that are first order in $N$. One can model the stochastic nature of the refractive index fluctuation by
\begin{equation}
\tilde{n}(\mathbf{x},z') = \int \exp[-\rmi 2\pi (\mathbf{a}\cdot\mathbf{x}+cz)] \chi(\mathbf{a},c) \left[ \frac{\Phi_n(\mathbf{a},c)}{\Delta^3} \right]^{1/2} \rmd^2 a\ \rmd c ,
\label{modeln}
\end{equation}
where $c$ is the longitudinal spatial frequency component, $\Delta$ is a dimension parameter on the frequency domain and $\chi(\mathbf{a},c)$ is a three-dimensional, normally distributed, complex-valued random function. Since $\tilde{n}$ is a real-valued function, it implies that $\chi^*(\mathbf{a},c)=\chi(-\mathbf{a},-c)$. Moreover, it is assumed that this random function is delta-correlated,
\begin{equation}
\langle \chi(\mathbf{a}_1,c_1) \chi^*(\mathbf{a}_2,c_2) \rangle = \Delta^3 \delta(\mathbf{a}_1-\mathbf{a}_2) \delta(c_1-c_2) .
\label{verwrand}
\end{equation}
The Kolmogorov power spectral density for the refractive index fluctuations is given by $\Phi_n = 0.033 (2\pi)^3 C_n^2 |\mathbf{k}|^{-11/3}$, where $C_n^2$ is the refractive index structure constant and $\mathbf{k}$ is a three-dimensional wave vector --- the Fourier domain variable for the refractive index of the medium ($\mathbf{k}=\{2\pi\mathbf{a},2\pi c\}$). The extra power of $(2\pi)^3$ comes from a different convention for the definition of the Fourier transform \cite{iperr}.

Next, we evaluate
\begin{eqnarray}
& \int_{z_0}^z \int_{z_0}^{z_1} \langle N(\mathbf{a},z_1) N(\mathbf{a}',z_2)\rangle\ \rmd z_2\ \rmd z_1  \nonumber \\
 = & \delta(\mathbf{a}+\mathbf{a}') \int \int_{z_0}^z \int_{z_0}^{z_1} \Phi_n(\mathbf{a},c_1) \exp[-\rmi 2\pi c_1 (z_1 - z_2)]\ \rmd z_2\ \rmd z_1\ \rmd c_1 ,
\end{eqnarray}
where we used (\ref{modeln}), (\ref{verwrand}) and the fact that $\Phi_n(\mathbf{a},c)$ is symmetric in all its arguments. Under the Markov approximation, one can replace $\Phi_n(\mathbf{a},c_1)\rightarrow\Phi_n(\mathbf{a},0)$, so that
\begin{equation}
\int_{z_0}^z \int_{z_0}^{z_1} \langle N(\mathbf{a},z_1) N(\mathbf{a}',z_2)\rangle\ \rmd z_2\ \rmd z_1
 = \frac{1}{2} (z-z_0) \delta(\mathbf{a}+\mathbf{a}') \Phi_n(\mathbf{a},0) .
\end{equation}
The expression in (\ref{eamm}) then becomes
\begin{equation}
\fl \int_{z_0}^z \int_{z_0}^{z_1} \langle M(\mathbf{a}_1,\mathbf{a}_2,z_1) M(\mathbf{a}_3,\mathbf{a}_4,z_2)\rangle\ \rmd z_2\ \rmd z_1
= \frac{dz}{2} k^2 \delta(\mathbf{a}_1-\mathbf{a}_2+\mathbf{a}_3-\mathbf{a}_4) \Phi_n(\mathbf{a}_1-\mathbf{a}_2,0) .
\label{eamms}
\end{equation}
where $dz=z-z_0$ and where we dropped a term that is second order in $dz$.

\subsection{The infinitesimal propagation functional equation}

With (\ref{eam0}) and (\ref{eamms}) substituted into (\ref{tweede1}), one can apply the limit $dz\rightarrow 0$. The resulting functional differential equation reads
\begin{eqnarray}
\partial_z W_{\hat{\rho}} =& \rmi \pi\lambda \int |\mathbf{a}|^2 \left[ \alpha^*(\mathbf{a}) \frac{\delta W_{\hat{\rho}}}{\delta\alpha^*(\mathbf{a})}
- \alpha(\mathbf{a}) \frac{\delta W_{\hat{\rho}}}{\delta\alpha(\mathbf{a})}\right]\ \rmd ^2 a \nonumber \\
& - \frac{1}{2} k^2 \Lambda \int \left[\alpha(\mathbf{a})\frac{\delta W_{\hat{\rho}}}{\delta\alpha(\mathbf{a})}
+ \alpha^*(\mathbf{a})\frac{\delta W_{\hat{\rho}}}{\delta\alpha^*(\mathbf{a})}\right]\ \rmd ^2 a \nonumber \\
& - \frac{1}{2} k^2 \int \left[\alpha(\mathbf{a}_1-\mathbf{a}_0)\alpha(\mathbf{a}_2+\mathbf{a}_0)\frac{\delta^2 W_{\hat{\rho}}}{\delta\alpha(\mathbf{a}_1)\delta\alpha(\mathbf{a}_2)} \right. \nonumber \\
& + \alpha^*(\mathbf{a}_1-\mathbf{a}_0)\alpha^*(\mathbf{a}_2+\mathbf{a}_0)\frac{\delta^2 W_{\hat{\rho}}}{\delta\alpha^*(\mathbf{a}_1)\delta\alpha^*(\mathbf{a}_2)} \nonumber \\
& \left. -2\alpha^*(\mathbf{a}_2-\mathbf{a}_0)\alpha(\mathbf{a}_1-\mathbf{a}_0)\frac{\delta^2 W_{\hat{\rho}}}{\delta\alpha(\mathbf{a}_1)\delta\alpha^*(\mathbf{a}_2)}\right]  \nonumber \\
& \times \Phi_n(\mathbf{a}_0,0)\ \rmd ^2 a_0\ \rmd ^2 a_1\ \rmd ^2 a_2 ,
\label{ipes}
\end{eqnarray}
where
\begin{equation}
\Lambda \equiv \int \Phi_n(\mathbf{a},0)\ \rmd ^2 a .
\end{equation}
The equation in (\ref{ipes}), which is our main result, represents the infinitesimal propagation functional equation (IPFE) for the evolution of an arbitrary multi-photon state, propagating through atmospheric turbulence under arbitrary conditions. It has the form of a Fokker-Planck equation for the Wigner functional of the state. The first term on the right is a drift term for free-space propagation, the second term is a drift term for the scintillation and the third is the diffusion term for the scintillation.

\section{\label{oplos}Solving the IPFE}

Apart from some trivial solutions, such as the vacuum and thermal states, no general solution has yet been found for the IPFE. Here we make a few observations and suggest approaches that can be used to find solutions for the IPFE.

\subsection{\label{fssol}Free-space solution}

In the case where there is no turbulence, one can use the first-order free-space equation given in (\ref{fsdv}). A solution for the free-space equation can be obtained by assuming a general form
\begin{eqnarray}
W_{\hat{\rho},fs}(z) = & \exp\left[-\alpha^*\diamond A(z)\diamond\alpha-\alpha\diamond B(z)\diamond\alpha-\alpha^*\diamond C(z)\diamond\alpha^* \right. \nonumber \\
& \left. +\alpha^*\diamond \beta(z)+\eta^*(z)\diamond\alpha\right] ,
\label{fsopl0}
\end{eqnarray}
where we ignored a possible normalization factor; $A(\mathbf{a},\mathbf{a}',z)$, $B(\mathbf{a},\mathbf{a}',z)$, and $C(\mathbf{a},\mathbf{a}',z)$ are unknown kernel (or covariance) functions; and $\beta(\mathbf{a},z)$ and $\eta(\mathbf{a},z)$ are unknown spectral functions.

Substituting (\ref{fsopl0}) into (\ref{fsdv}), and separating the resulting equation according to the different combinations of contraction with $\alpha$ and $\alpha^*$, we obtain five separate differential equations for the three kernels and the two spectral functions
\begin{equation}
\eqalign{
\partial_z A(\mathbf{a}_1,\mathbf{a}_2,z) = \rmi\pi\lambda \left(|\mathbf{a}_1|^2-|\mathbf{a}_2|^2\right) A(\mathbf{a}_1,\mathbf{a}_2,z) \\
\partial_z B(\mathbf{a}_1,\mathbf{a}_2,z) = -\rmi\pi\lambda\left(|\mathbf{a}_1|^2+|\mathbf{a}_2|^2\right) B(\mathbf{a}_1,\mathbf{a}_2,z) \\
\partial_z C(\mathbf{a}_1,\mathbf{a}_2,z) = \rmi\pi\lambda\left(|\mathbf{a}_1|^2+|\mathbf{a}_2|^2\right) C(\mathbf{a}_1,\mathbf{a}_2,z) \\
\partial_z \beta(\mathbf{a},z) = \rmi\pi\lambda |\mathbf{a}|^2 \beta(\mathbf{a},z) \\
\partial_z \eta^*(\mathbf{a},z) = -\rmi\pi\lambda |\mathbf{a}|^2 \eta^*(\mathbf{a},z) .}
\end{equation}
Their solutions are
\begin{equation}
\eqalign{
A(\mathbf{a}_1,\mathbf{a}_2,z) = A_0(\mathbf{a}_1,\mathbf{a}_2) \exp\left[\rmi \pi\lambda z\left(|\mathbf{a}_1|^2-|\mathbf{a}_2|^2\right)\right] \\
B(\mathbf{a}_1,\mathbf{a}_2,z) = B_0(\mathbf{a}_1,\mathbf{a}_2) \exp\left[-\rmi \pi\lambda z\left(|\mathbf{a}_1|^2+|\mathbf{a}_2|^2\right)\right] \\
C(\mathbf{a}_1,\mathbf{a}_2,z) = C_0(\mathbf{a}_1,\mathbf{a}_2) \exp\left[\rmi \pi\lambda z\left(|\mathbf{a}_1|^2+|\mathbf{a}_2|^2\right)\right] \\
\beta(\mathbf{a},z) = \beta_0(\mathbf{a}) \exp\left(\rmi\pi\lambda z|\mathbf{a}|^2\right) \\
\eta^*(\mathbf{a},z) = \eta_0^*(\mathbf{a}) \exp\left(-\rmi\pi\lambda z|\mathbf{a}|^2\right) ,}
\end{equation}
where $A_0(\mathbf{a}_1,\mathbf{a}_2)$, $B_0(\mathbf{a}_1,\mathbf{a}_2)$, and $C_0(\mathbf{a}_1,\mathbf{a}_2)$ are the initial kernels (or covariance functions) and $\beta_0(\mathbf{a})$, and $\eta_0^*(\mathbf{a})$ are the initial spectral functions. Hence, the free-space solution is
\begin{eqnarray}
\fl W_{\hat{\rho},fs}(z) =& \exp\left(-\alpha^*\diamond\mathbf{1}P\diamond A_0\diamond\mathbf{1}P^*\diamond\alpha
-\alpha\diamond\mathbf{1}P^*\diamond B_0\diamond\mathbf{1}P^*\diamond\alpha \right. \nonumber \\
& \left. -\alpha^*\diamond\mathbf{1}P\diamond C_0\diamond\mathbf{1}P\diamond\alpha^*
+\alpha^*\diamond\mathbf{1}P\diamond\beta_0+\eta_0^*\diamond\mathbf{1}P^*\diamond\alpha\right) ,
\label{fsopl}
\end{eqnarray}
where
\begin{equation}
\mathbf{1}P = \exp\left(\rmi\pi\lambda z|\mathbf{a}|^2\right)\delta(\mathbf{a}-\mathbf{a}') .
\label{peedef}
\end{equation}
Another way to view it, is that free-space propagation implies a transformation of the $\alpha$'s, given by
\begin{equation}
\eqalign{
\alpha \rightarrow \exp\left(-\rmi\pi\lambda z|\mathbf{a}|^2\right)\alpha \\
\alpha^* \rightarrow \exp\left(\rmi\pi\lambda z|\mathbf{a}|^2\right)\alpha^* .}
\end{equation}
As such, the result corresponds to simple free-space propagation in a way that one would expect.

\subsection{Gaussian solutions}

While the solution without turbulence provides an important cross-check to see that the first-order equation is consistent with our expectations, it does not give any indication of the validity of the IPFE. Including the effect of the turbulent medium, one may consider solutions for the IPFE that have the form of a Gaussian Wigner functional. Although, we find only trivial solutions, they do serve as a cross-check for the IPFE.

Consider, for instance the simplest case where the solution has the form
\begin{equation}
W_{\hat{\rho}}(z) = \exp\left[ -\alpha^*\diamond A(z)\diamond\alpha \right] ,
\end{equation}
with $A(\mathbf{a},\mathbf{a}',z)$ being an unknown $z$-dependent kernel function. When we substitute this ansatz into (\ref{ipes}), evaluate all the functional derivatives and the $z$-derivative, and remove a factor of the original Wigner functional $W_{\hat{\rho}}(z)$ on both sides, we obtain
\begin{eqnarray}
\fl - \alpha^*\diamond \partial_z A(z)\diamond\alpha =& -\rmi \pi\lambda \int \left(|\mathbf{a}|^2-|\mathbf{a}'|^2\right)
 \alpha^*(\mathbf{a}) A(\mathbf{a},\mathbf{a}',z)\alpha(\mathbf{a}')\ \rmd ^2 a\ \rmd ^2 a' \nonumber \\
& + k^2 \Lambda \int \alpha^*(\mathbf{a}) A(\mathbf{a},\mathbf{a}',z) \alpha(\mathbf{a}')\ \rmd ^2 a\ \rmd ^2 a' \nonumber \\
& - k^2 \int \alpha^*(\mathbf{a}_2-\mathbf{a}_0)A(\mathbf{a}_2,\mathbf{a}_1,z)\alpha(\mathbf{a}_1-\mathbf{a}_0) \Phi_n(\mathbf{a}_0,0)\ \rmd ^2 a_0\ \rmd ^2 a_1\ \rmd ^2 a_2 \nonumber \\
& + \frac{1}{2} k^2 \int \left[\alpha^*(\mathbf{a}_3)A(\mathbf{a}_3,\mathbf{a}_1,z)\alpha(\mathbf{a}_1-\mathbf{a}_0)\alpha^*(\mathbf{a}_4)A(\mathbf{a}_4,\mathbf{a}_2,z)\alpha(\mathbf{a}_2+\mathbf{a}_0) \right. \nonumber \\
& +\alpha^*(\mathbf{a}_1-\mathbf{a}_0)A(\mathbf{a}_1,\mathbf{a}_3,z)\alpha(\mathbf{a}_3) \alpha^*(\mathbf{a}_2+\mathbf{a}_0)A(\mathbf{a}_2,\mathbf{a}_4,z)\alpha(\mathbf{a}_4) \nonumber \\
& \left. -2\alpha^*(\mathbf{a}_2-\mathbf{a}_0)A(\mathbf{a}_2,\mathbf{a}_4,z)\alpha(\mathbf{a}_4) \alpha^*(\mathbf{a}_3) A(\mathbf{a}_3,\mathbf{a}_1,z)\alpha(\mathbf{a}_1-\mathbf{a}_0)\right] \nonumber \\
& \times \Phi_n(\mathbf{a}_0,0)\ \rmd ^2 a_0\ \rmd ^2 a_1\ \rmd ^2 a_2\ \rmd ^2 a_3\ \rmd ^2 a_4 .
\label{isogauss}
\end{eqnarray}
We see that the equation does not contain only second-order terms (with single factors of both $\alpha$ and $\alpha^*$), but also fourth-order terms (with two factors of $\alpha$ and $\alpha^*$ each). As a result, the equation separates into two equations that must be satisfied separately. The second-order equation is
\begin{eqnarray}
\partial_z A(\mathbf{a},\mathbf{a}',z) =& \rmi \pi\lambda \left(|\mathbf{a}|^2-|\mathbf{a}'|^2\right) A(\mathbf{a},\mathbf{a}',z) - k^2 \Lambda A(\mathbf{a},\mathbf{a}',z)\nonumber \\
& + k^2 \int A(\mathbf{a}+\mathbf{a}_0,\mathbf{a}'+\mathbf{a}_0,z) \Phi_n(\mathbf{a}_0,0)\ \rmd ^2 a_0 ,
\end{eqnarray}
and the fourth-order equation is
\begin{eqnarray}
0 =& \int \left[ A(\mathbf{a}_1,\mathbf{a}_2+\mathbf{a}_0,z) A(\mathbf{a}_3,\mathbf{a}_4-\mathbf{a}_0,z) \right. \nonumber \\
& + A(\mathbf{a}_1+\mathbf{a}_0,\mathbf{a}_2,z) A(\mathbf{a}_3-\mathbf{a}_0,\mathbf{a}_4,z)   \nonumber \\
& \left. -2A(\mathbf{a}_1+\mathbf{a}_0,\mathbf{a}_2,z) A(\mathbf{a}_3,\mathbf{a}_4+\mathbf{a}_0,z)\right] \nonumber \\
& \times \Phi_n(\mathbf{a}_0,0)\ \rmd ^2 a_0\ \rmd ^2 a_1\ \rmd ^2 a_2\ \rmd ^2 a_3\ \rmd ^2 a_4 ,
\end{eqnarray}
where we shifted the $\mathbf{a}_0$ into the arguments of the kernels so that we can removing all the common factors and the identical contractions with the $\alpha$'s and $\alpha^*$'s. Expressing the kernels in terms of sums and differences of their arguments, we have
\begin{eqnarray}
0 =& \int \left[ A'(\mathbf{a}_1+\mathbf{a}_2+\mathbf{a}_0,\mathbf{a}_1-\mathbf{a}_2-\mathbf{a}_0,z)  A'(\mathbf{a}_3+\mathbf{a}_4-\mathbf{a}_0,\mathbf{a}_3-\mathbf{a}_4+\mathbf{a}_0,z) \right. \nonumber \\
& +  A'(\mathbf{a}_1+\mathbf{a}_2+\mathbf{a}_0,\mathbf{a}_1-\mathbf{a}_2+\mathbf{a}_0,z)  A'(\mathbf{a}_3+\mathbf{a}_4-\mathbf{a}_0,\mathbf{a}_3-\mathbf{a}_4-\mathbf{a}_0,z)  \nonumber \\
& \left. -2 A'(\mathbf{a}_1+\mathbf{a}_2+\mathbf{a}_0,\mathbf{a}_1-\mathbf{a}_2+\mathbf{a}_0,z)  A'(\mathbf{a}_3+\mathbf{a}_4+\mathbf{a}_0,\mathbf{a}_3-\mathbf{a}_4-\mathbf{a}_0,z)\right]\nonumber \\
& \times \Phi_n(\mathbf{a}_0,0)\ \rmd ^2 a_0 .
\end{eqnarray}
It now follows that, if the kernel only depend on the difference in arguments $\mathbf{a}_d$, then the sign of $\mathbf{a}_0$ in the first term can be changed (since $\Phi_n$ is symmetric) and all the terms would cancel. However, this condition would also affect the second-order equation, causing the terms associated with the scintillation to cancel, leaving only the free-space propagation effect. Hence, we end up with
\begin{eqnarray}
\partial_z A(\mathbf{a}-\mathbf{a}',z) =& \rmi \pi\lambda \left(|\mathbf{a}|^2-|\mathbf{a}'|^2\right) A(\mathbf{a}-\mathbf{a}',z) ,
\end{eqnarray}
The only solutions are those where $A(\mathbf{a}-\mathbf{a}',z) \propto \delta(\mathbf{a}-\mathbf{a}')$. These include the vacuum state and the thermal states.

As a second attempt to solve (\ref{ipes}), we introduce some shift terms
\begin{equation}
W_{\hat{\rho}}(z) = \exp\left[ -\alpha^*\diamond A(z)\diamond\alpha-\alpha^*\diamond \beta(z)-\eta^*(z)\diamond\alpha \right] .
\end{equation}
However, after substituting it into (\ref{fsdv}) and separating the result into separate equations according to the contractions with $\alpha$ and $\alpha^*$, one obtains differential equations for the two complex functions given by
\begin{equation}
\eqalign{
\partial_z \beta(\mathbf{a},z) = \rmi \pi\lambda |\mathbf{a}|^2 \beta(\mathbf{a},z) - k^2 \Lambda \beta(\mathbf{a},z) \\
\partial_z \eta^*(\mathbf{a},z) = -\rmi \pi\lambda |\mathbf{a}|^2 \eta^*(\mathbf{a},z) - k^2 \Lambda \eta^*(\mathbf{a},z) .}
\label{ordeen}
\end{equation}
Their solutions are
\begin{equation}
\eqalign{
\beta(\mathbf{a},z) = \exp(\rmi \pi\lambda z |\mathbf{a}|^2 - k^2 \Lambda z) , \\
\eta^*(\mathbf{a},z) = \exp(-\rmi \pi\lambda z |\mathbf{a}|^2 - k^2 \Lambda z) .}
\label{ordeenopl}
\end{equation}
However, since $\Lambda$ is a divergent distance, the result is that $\beta(z)=\eta^*(z)=0$ for $z>0$. In addition, there are combinations of $\alpha$'s and $\alpha^*$'s produced on the right-hand side of the equation that are not represented by any term on the left-hand side. In all these cases, the requirement $\beta(z)=\eta^*(z)=0$ is enforced. It thus leads back to the unshifted previous result.

One can proceed to search for a Gaussian solution having an even more complicated expression in the exponent. However, the resulting equation would produce several separate second-order equations and fourth-order equations, which would be difficult to solve, apart from the trivial solution. In the end, the only solutions for the IPFE that can be readily found by assuming a Gaussian form for the Wigner functional are those with isotropic arguments, centered at the origin, such as the vacuum state and the thermal states. Although they do not represent general solutions for the IPFE, they serve as an important cross-check, because it makes sense that these states would not evolve during propagation through a turbulent atmosphere.

\subsection{Characteristic functional}

Another way to search for a general solution for the IPFE in (\ref{ipes}), is to transform the Wigner functional to its corresponding characteristic functional. The hope is that the result would provide insight into the required form for the solution.

The Wigner functional is given in terms of the characteristic functional by
\begin{equation}
W[\alpha] = \int \exp(\alpha^*\diamond\eta-\eta^*\diamond\alpha) \chi[\eta]\ \mathcal{D}[\eta] ,
\label{wigchar}
\end{equation}
and the inverse process is given by
\begin{equation}
\chi[\eta] = \int W[\alpha] \exp(\eta^*\diamond\alpha-\alpha^*\diamond\eta)\ \mathcal{D}[\alpha] .
\label{charwig}
\end{equation}
First, we substitute (\ref{wigchar}) into (\ref{ipes}), with $\eta\rightarrow\eta_0$, and then we apply the inverse process given in (\ref{charwig}). The result has the form
\begin{equation}
\fl \partial_z \chi[\eta] = \int \exp(\alpha^*\diamond\eta_0-\eta_0^*\diamond\alpha+\eta^*\diamond\alpha-\alpha^*\diamond\eta) K[\alpha,\eta_0] \chi[\eta_0](z)\ \mathcal{D}[\eta_0,\alpha] ,
\label{ch0}
\end{equation}
where
\begin{eqnarray}
K[\alpha,\eta_0] =& \rmi\pi\lambda \int |\mathbf{a}|^2\left[\alpha^*(\mathbf{a})\eta_0(\mathbf{a})+\eta_0^*(\mathbf{a})\alpha(\mathbf{a})\right]\ \rmd ^2 a \nonumber \\
& - \frac{1}{2} k^2 \Lambda \int \left[\alpha^*(\mathbf{a})\eta_0(\mathbf{a})-\eta_0^*(\mathbf{a})\alpha(\mathbf{a})\right]\ \rmd ^2 a \nonumber \\
& - \frac{1}{2} k^2 \int \left[ \alpha(\mathbf{a}_2)\alpha(\mathbf{a}_1)\eta_0^*(\mathbf{a}_1+\mathbf{a}_0)\eta_0^*(\mathbf{a}_2-\mathbf{a}_0) \right. \nonumber \\
& + \alpha^*(\mathbf{a}_1)\alpha^*(\mathbf{a}_2)\eta_0(\mathbf{a}_1+\mathbf{a}_0)\eta_0(\mathbf{a}_2-\mathbf{a}_0) \nonumber \\
& \left. + 2 \alpha^*(\mathbf{a}_1)\alpha(\mathbf{a}_2)\eta_0(\mathbf{a}_1-\mathbf{a}_0)\eta_0^*(\mathbf{a}_2-\mathbf{a}_0)\right] \nonumber \\
& \times \Phi_n(\mathbf{a}_0,0)\ \rmd ^2 a_0\ \rmd ^2 a_1\ \rmd ^2 a_2 .
\label{defk}
\end{eqnarray}
Note that the positive sign in the second last line comes about due to the different signs of the terms in the exponent in (\ref{ch0}).

For the subsequent calculation, we construct $K[\alpha,\eta_0]$ with the aid of functional derivatives applied to a source functional of the form
\begin{eqnarray}
\mathcal{S}[\eta_0,\eta,\nu,\mu] = & \int \exp\left(\alpha^*\diamond\eta_0-\eta_0^*\diamond\alpha+\eta^*\diamond\alpha-\alpha^*\diamond\eta \right. \nonumber \\
& \left. +\alpha^*\diamond\nu-\nu^*\diamond\alpha+\mu^*\diamond\eta_0-\eta_0^*\diamond\mu\right)\ \mathcal{D}[\alpha] ,
\label{bronf}
\end{eqnarray}
where $\nu$ and $\mu$ are auxiliary functions. One can define a {\em construction operator} that would produce $K[\alpha,\eta_0]$ from the source functional by replacing
\begin{eqnarray}
\eqalign{
\alpha \rightarrow & -\frac{\delta}{\delta\nu^*} \\
\alpha^* \rightarrow & \frac{\delta}{\delta\nu} \\
\eta_0 \rightarrow & \frac{\delta}{\delta\mu^*} \\
\eta_0^* \rightarrow & -\frac{\delta}{\delta\mu} , }
\end{eqnarray}
in (\ref{defk}). The result is a functional differential operator
\begin{eqnarray}
\mathcal{K} = & \rmi\pi\lambda \int |\mathbf{a}|^2\left[\frac{\delta}{\delta\nu(\mathbf{a})}\frac{\delta}{\delta\mu^*(\mathbf{a})}
+\frac{\delta}{\delta\mu(\mathbf{a})}\frac{\delta}{\delta\nu^*(\mathbf{a})}\right]\ {\rm d}^2 a \nonumber \\
& - \frac{1}{2} k^2 \Lambda \int \left[\frac{\delta}{\delta\nu(\mathbf{a})}\frac{\delta}{\delta\mu^*(\mathbf{a})}
 -\frac{\delta}{\delta\mu(\mathbf{a})}\frac{\delta}{\delta\nu^*(\mathbf{a})}\right]\ {\rm d}^2 a \nonumber \\
& - \frac{1}{2} k^2 \int \left[ \frac{\delta}{\delta\mu(\mathbf{a}_1+\mathbf{a}_0)} \frac{\delta}{\delta\mu(\mathbf{a}_2-\mathbf{a}_0)}
\frac{\delta}{\delta\nu^*(\mathbf{a}_2)} \frac{\delta}{\delta\nu^*(\mathbf{a}_1)} \right. \nonumber \\
& + \frac{\delta}{\delta\nu(\mathbf{a}_1)} \frac{\delta}{\delta\nu(\mathbf{a}_2)}
\frac{\delta}{\delta\mu^*(\mathbf{a}_1+\mathbf{a}_0)} \frac{\delta}{\delta\mu^*(\mathbf{a}_2-\mathbf{a}_0)} \nonumber \\
& \left. + 2 \frac{\delta}{\delta\nu(\mathbf{a}_1)} \frac{\delta}{\delta\mu^*(\mathbf{a}_1-\mathbf{a}_0)}
\frac{\delta}{\delta\mu(\mathbf{a}_2-\mathbf{a}_0)} \frac{\delta}{\delta\nu^*(\mathbf{a}_2)}\right]\nonumber \\
& \times \Phi_n(\mathbf{a}_0,0)\ {\rm d}^2 a_0\ {\rm d}^2 a_1\ {\rm d}^2 a_2 .
\label{defkon}
\end{eqnarray}
The equation for the characteristic function is then represented by
\begin{equation}
\partial_z \chi[\eta] = \left. \mathcal{K}\left\{ \int \mathcal{S}[\eta_0,\eta,\nu,\mu] \chi[\eta_0]\ \mathcal{D}[\eta_0] \right\} \right|_{\nu=\mu=0} .
\label{ch4}
\end{equation}

The functional integration over $\alpha$ in (\ref{bronf}) produces a Dirac delta functional together with an exponential
\begin{equation}
\mathcal{S}[\eta_0,\eta,\nu,\mu] = \delta\left[\eta_0-\eta+\nu,\eta_0^*-\eta^*+\nu^*\right] \exp(\mu^*\diamond\eta_0-\eta_0^*\diamond\mu) .
\end{equation}
When it is substituted into (\ref{ch4}), the functional integration over $\eta_0$ causes the Dirac delta functional to replace the argument of the characteristic functional with expressions containing the auxiliary functions. The resulting equation has the form
\begin{equation}
\fl \partial_z \chi[\eta,\eta^*] = \left. \mathcal{K}\left\{ \chi[\eta-\nu,\eta^*-\nu^*]  \exp[\mu^*\diamond(\eta-\nu)-(\eta^*-\nu^*)\diamond\mu] \right\} \right|_{\nu=\mu=0} .
\label{ch5}
\end{equation}
One can see that the operation of $\mathcal{K}$ on the shifted characteristic functional would lead to a functional differential equation.

We substitute (\ref{defkon}) into (\ref{ch5}), apply the functional derivatives on the characteristic functional and set $\nu=\mu=0$. The resulting equation reads
\begin{eqnarray}
\partial_z \chi[\eta,\eta^*] = & \rmi\pi\lambda \int |\mathbf{a}|^2 \left\{\eta^*(\mathbf{a}) \frac{\delta\chi[\eta,\eta^*]}{\delta\eta^*(\mathbf{a})}
- \eta(\mathbf{a}) \frac{\delta\chi[\eta,\eta^*]}{\delta\eta(\mathbf{a})} \right\}\ {\rm d}^2 a \nonumber \\
& - \frac{1}{2}k^2 \Lambda \int \left\{ \eta(\mathbf{a}) \frac{\delta\chi[\eta,\eta^*]}{\delta\eta(\mathbf{a})}
+ \eta^*(\mathbf{a}) \frac{\delta\chi[\eta,\eta^*]}{\delta\eta^*(\mathbf{a})} \right\}\ {\rm d}^2 a \nonumber \\
& - \frac{1}{2}k^2 \int \left\{\eta(\mathbf{a}_1+\mathbf{a}_0)\eta(\mathbf{a}_2-\mathbf{a}_0)\frac{\delta^2\chi[\eta,\eta^*]}{\delta\eta(\mathbf{a}_1)\delta\eta(\mathbf{a}_2)} \right. \nonumber \\
& +\eta^*(\mathbf{a}_1+\mathbf{a}_0)\eta^*(\mathbf{a}_2-\mathbf{a}_0)\frac{\delta^2\chi[\eta,\eta^*]}{\delta\eta^*(\mathbf{a}_1)\delta\eta^*(\mathbf{a}_2)} \nonumber \\
& \left. -2\eta(\mathbf{a}_1-\mathbf{a}_0)\eta^*(\mathbf{a}_2-\mathbf{a}_0)\frac{\delta^2\chi[\eta,\eta^*]}{\delta\eta(\mathbf{a}_1)\delta\eta^*(\mathbf{a}_2)} \right\} \nonumber \\
& \times \Phi_n(\mathbf{a}_0,0)\ {\rm d}^2 a_0\ {\rm d}^2 a_1\ {\rm d}^2 a_2 ,
\label{cffe}
\end{eqnarray}
It has the form of a Fokker-Planck equation --- the same form as the IPFE for the Wigner functional in (\ref{ipes}). As such, the characteristic functional approach does not bring us much nearer to a solution, but it does indicate that the solution of the IPFE and its functional Fourier transform satisfy the same functional differential equation. It thus suggests that some solutions of the IPFE may be eigenstates of the functional Fourier transform. This is indeed true for the Wigner functional of the vacuum state. On the other hand, the Wigner functionals of the thermal states are not equal to their characteristic functionals, but they all have the same form and are therefore (trivial) solutions of these functional differential equations.


\subsection{Fixed photon-number solutions}

Although the functional Fokker-Planck equation in (\ref{ipes}) involves arbitrary multi-photon states, it must also incorporate situations where the state has a fixed number of photons. One can argue that the photon number must remain the same during propagation, because the process from which (\ref{ipes}) is derived does not include a loss-mechanism. Therefore, if the states start out with exactly $n$ photons, then they must contain $n$ photons all the time. One can therefore apply the process to a general Fock state, in which the spectrum and its complex conjugate combines to form the density matrix for a state with a fixed number of photons.

For this purpose, we can use a generating functional for the Wigner functionals of fixed-spectrum Fock states. The one-parameter version of this generating functional (\ref{fockwig2}), is derived in \ref{genfock}.\footnote{The two-parameter version of this generating functional is derived in \cite{wigfunk}.} Since this generating functional has the form of an isotropic Gaussian, we can use the result obtained in (\ref{isogauss}), but without removing the factor of the original Wigner functional. One can then evaluate the number of derivatives with respect to the auxiliary variable equal to the number of photons in the state, before setting the auxiliary parameter to zero.

For the single-photon solution, we evaluate one derivative with respect the auxiliary variable, set it to zero and remove the remaining Gaussian factor to obtain
\begin{eqnarray}
\alpha^*\diamond\partial_z \rho(z)\diamond\alpha = & i \pi \lambda \left( |\mathbf{a}_1|^2 - |\mathbf{a}_2|^2 \right) \alpha^*\diamond \rho(z)\diamond\alpha \nonumber \\
& - k^2 \int \Phi_0(\mathbf{u}) \left[\alpha^*(\mathbf{a}_1)\rho(\mathbf{a}_1,\mathbf{a}_2,z)\alpha(\mathbf{a}_2) \right. \nonumber \\
& \left. - \alpha^*(\mathbf{a}_1)\rho(\mathbf{a}_1-\mathbf{u},\mathbf{a}_2-\mathbf{u},z)\alpha(\mathbf{a}_2) \right]\ {\rm d}^2 u\ {\rm d}^2 a_1\ {\rm d}^2 a_2 ,
\end{eqnarray}
where $\rho(z)$ is the density matrix in the spatial frequency basis for the single-photon state. The remaining $\alpha$'s and $\alpha^*$'s can be removed by functional derivatives. The result would then be appropriately symmetrized. It is equivalent to the single-photon infinitesimal propagation equation (IPE), which has been solved before, using the quadratic structure function approximation \cite{notrunc}.

For the bi-photon solution, we evaluate two derivatives with respect the auxiliary variable, set it to zero and remove the remaining Gaussian factor. In this case, one obtains the two-photon equation together with the single-photon equation. After removing the single-photon equation and the $\alpha$'s and $\alpha^*$'s with functional derivatives, we obtain the symmetrized equation
\begin{eqnarray}
\partial_z \rho(z) & = & i\pi \lambda \left( |\mathbf{a}_1|^2 - |\mathbf{a}_2|^2 + |\mathbf{a}_3|^2 - |\mathbf{a}_4|^2 \right) F(z)
- k^2 \int \Phi_0(\mathbf{u}) \left[ 2 F(z) \right. \nonumber \\
& & - F(\mathbf{a}_1-\mathbf{u},\mathbf{a}_2-\mathbf{u},\mathbf{a}_3,\mathbf{a}_4,z)
- F(\mathbf{a}_1,\mathbf{a}_2,\mathbf{a}_3-\mathbf{u},\mathbf{a}_4-\mathbf{u},z) \nonumber \\
& & - F(\mathbf{a}_1-\mathbf{u},\mathbf{a}_2,\mathbf{a}_3,\mathbf{a}_4-\mathbf{u},z)
- F(\mathbf{a}_1,\mathbf{a}_2-\mathbf{u},\mathbf{a}_3-\mathbf{u},\mathbf{a}_4,z) \nonumber \\
& & \left. + F(\mathbf{a}_1-\mathbf{u},\mathbf{a}_2,\mathbf{a}_3+\mathbf{u},\mathbf{a}_4,z)
+ F(\mathbf{a}_1,\mathbf{a}_2-\mathbf{u},\mathbf{a}_3,\mathbf{a}_4+\mathbf{u},z) \right] \nonumber \\
& & \times {\rm d}^2 u .
\end{eqnarray}
It is the same as the bi-photon IPE, for which a solution has previously been found under the quadratic structure function approximation \cite{kormed}.

These two cases show that the multi-photon Fokker-Planck equation in (\ref{ipes}) is consistent with previous results. An equation for any fixed photon-number state can be obtained in this way. However, they become progressively more difficult to solve.


\subsection{Polynomial solution}

Due to the requirement for normalizability of the state, it is not expected that the solution of the IPFE would be a polynomial of finite order. Such polynomials are unbounded and therefore not normalizable. However, any normalizable functional can be represented as a polynomial of infinite order --- a transcendental polynomial.

Here, we express the solution in the form of such a transcendental polynomial and substitute it into the IPFE. The result is an infinite set of uncoupled differential equations. If all these differential equations can somehow be solved, one would have a general solution for the IPFE.

The expression of the transcendental functional polynomial is
\begin{eqnarray}
W_{\rm pol}[\alpha,\alpha^*](z) &=& \sum_{m,n=0}^{\infty} \int \alpha^*(\mathbf{a}_1) ... \alpha^*(\mathbf{a}_m)
 H_{m,n}(\mathbf{a}_1, ..., \mathbf{a}_m,\mathbf{a}_1', ..., \mathbf{a}_n',z) \nonumber \\
&& \times \alpha(\mathbf{a}_1') ... \alpha(\mathbf{a}_n')\ \rmd ^2 a_1 ... \rmd ^2 a_m\ \rmd ^2 a_1' ... \rmd ^2 a_n'  \nonumber \\
  &=& \sum_{m,n=0}^{\infty} \left( \prod^m \alpha^* \right) \odot H_{m,n}(z) \odot \left( \prod^n \alpha \right) ,
\label{pols0}
\end{eqnarray}
where $\odot$ represents multiple contractions, connecting the kernels $H_{m,n}$ to all the $\alpha^*$'s and $\alpha$'s. Note that the kernels are symmetric with respect to any perturbation among $\{\mathbf{a}_1 ... \mathbf{a}_m\}$ or among $\{\mathbf{a}_1' ... \mathbf{a}_n'\}$.

When we substitute the expression for the functional polynomial into the IPFE in (\ref{ipes}), we get an expression that can be separated into separate equations for every order. These equations are of the form
\begin{eqnarray}
\fl \partial_z H_{m,n}(z) =& \rmi \pi\lambda \left( m |\mathbf{a}_1|^2 - n |\mathbf{a}_1'|^2 \right) H_{m,n}(\mathbf{a}_1, ..., \mathbf{a}_m,\mathbf{a}_1', ..., \mathbf{a}_n',z) \nonumber \\
& - \frac{1}{2} k^2 \Lambda (m+n) H_{m,n}(\mathbf{a}_1, ..., \mathbf{a}_m,\mathbf{a}_1', ..., \mathbf{a}_n',z) \nonumber \\
& - \frac{1}{2} k^2 \int \left[ m(m-1) H_{m,n}(\mathbf{a}_1+\mathbf{a}_0,\mathbf{a}_2-\mathbf{a}_0, ..., \mathbf{a}_m,\mathbf{a}_1', ..., \mathbf{a}_n',z) \right. \nonumber \\
& + n(n-1) H_{m,n}(\mathbf{a}_1, ..., \mathbf{a}_m,\mathbf{a}_1'+\mathbf{a}_0,\mathbf{a}_2'-\mathbf{a}_0, ..., \mathbf{a}_n',z) \nonumber \\
& \left. - 2 m n H_{m,n}(\mathbf{a}_1+\mathbf{a}_0, ..., \mathbf{a}_m,\mathbf{a}_1'+\mathbf{a}_0, ..., \mathbf{a}_n',z) \right] \Phi_n(\mathbf{a}_0,0)\ \rmd ^2 a_0 .
\label{genkernvg}
\end{eqnarray}
Due to the integer factors, which are produced by the functional derivatives, the equations simplify for lower values of $m$ and $n$. For $m\geq 2$ and $n\geq 2$ the equations are of the general form, given in (\ref{genkernvg}). We note that the solution for $H_{n,m}(z)$ is always the complex conjugate of the solution for $H_{m,n}(z)$.

The first few equations in the sequence are
\begin{equation}
\fl \eqalign{
\partial_z H_{0,0}(z) & = 0 , \\
\partial_z H_{1,0}(\mathbf{a}_1,z) & = \rmi \pi\lambda |\mathbf{a}_1|^2 H_{1,0}(\mathbf{a}_1,z) - \frac{1}{2}k^2 \Lambda H_{1,0}(\mathbf{a}_1,z) \\
\partial_z H_{0,1}(\mathbf{a}_1',z) & = -\rmi \pi\lambda |\mathbf{a}_1'|^2 H_{0,1}(\mathbf{a}_1',z) - \frac{1}{2}k^2 \Lambda H_{0,1}(\mathbf{a}_1',z) \\
\partial_z H_{1,1}(\mathbf{a}_1,\mathbf{a}_1',z) & = \rmi \pi\lambda \left[|\mathbf{a}_1|^2 - |\mathbf{a}_1'|^2\right] H_{1,1}(\mathbf{a}_1,\mathbf{a}_1',z)
 - k^2 \Lambda H_{1,1}(\mathbf{a}_1,\mathbf{a}_1',z) \\
 & ~~~ + k^2 \int H_{1,1}(\mathbf{a}_1+\mathbf{a}_0,\mathbf{a}_1'+\mathbf{a}_0,z) \Phi_n(\mathbf{a}_0,0)\ \rmd ^2 a_0} .
\label{eenvvgs}
\end{equation}
The first equation in (\ref{eenvvgs}) implies that $H_{0,0}$ is a constant independent of $z$. It can thus act as a normalization constant for the state. The solutions for $H_{1,0}(\mathbf{a}_1,z)$ and its complex conjugate are
\begin{equation}
\eqalign{
H_{1,0}(\mathbf{a}_1,z) = B_{1,0}(\mathbf{a}_1) \exp\left(\rmi \pi\lambda z |\mathbf{a}_1|^2 - \frac{1}{2} k^2 \Lambda z\right) \\
H_{0,1}(\mathbf{a}_1',z) = B_{0,1}(\mathbf{a}_1') \exp\left(-\rmi \pi\lambda z |\mathbf{a}_1'|^2 - \frac{1}{2} k^2 \Lambda z\right) ,}
\end{equation}
where $B_{1,0}(\mathbf{a}_1)$ and $B_{0,1}(\mathbf{a}_1')$ are the initial spectral functions at $z=0$. These solutions decay to zero on a scale equivalent to $z\sim 1/(k^2 \Lambda)$. The equation for $H_{1,1}(\mathbf{a}_1,\mathbf{a}_1',z)$ is equivalent to the IPE for a single-photon state, which has been solved in \cite{notrunc} under the quadratic structure function approximation. The solution is in the form of an integral
\begin{equation}
H_{1,1}(\mathbf{a}_1,\mathbf{a}_1',z) = \int B_{1,1}(\mathbf{a}_1-\mathbf{u},\mathbf{a}_1'-\mathbf{u}) L(\mathbf{a}_1,\mathbf{a}_1',\mathbf{u},z)\ \rmd ^2 u ,
\label{oplos11}
\end{equation}
where $L(\mathbf{a}_1,\mathbf{a}_1',\mathbf{u},z)$ is a kernel function given in \cite{notrunc} for the quadratic structure function approximation.

All the remaining equations in the sequence of equations given in (\ref{genkernvg}) still need to be solved. It is not at this time certain that a generic solution can be found for arbitrary orders in (\ref{genkernvg}). If such generic solutions can be found, it is still not certain that such solutions would produce results that can be summed into a closed form to serve as a general solution for the functional differential equation.

\section{Conclusion}
\label{concl}

The evolution of an arbitrary photonic quantum state propagating through a turbulent atmosphere is considered. An infinitesimal propagation approach is used to obtain a Fokker-Planck equation for the Wigner functional of the state --- the IPFE. Apart from the $z$-derivatives, all the other derivatives in the equation are functional derivatives.

We do not provide a general solution for the IPFE. However, we show that without turbulence, the solution is in the form as expected for free-space propagation without turbulence. Considering the possibility for a solution in the form of a Gaussian functional, we find that the only solutions of this form are the thermal states, which include the vacuum state.

The use of the characteristic functional to obtain a general solution is also considered. It leads to another functional differential equation for the characteristic functional that has the same form as the IPFE. One can therefore concluded that a general solution of the IPFE would to a set of functional that are mapped back onto itself by the functional Fourier transform.

We show that the functional Fokker-Planck equation is consistent with previous solutions that were obtained for single-photon and bi-photon states. It follows from the observation that the Fokker-Planck equation maintains photon-number, because it does not involve a loss mechanism.

Finally, we propose a general approach to find a solution, expressing it as an infinite order polynomial functional, which can then be decoupled into an infinite set of uncoupled differential equations. It is not sure whether an approach exists with which all such equations can be solved.

\section*{Acknowledgement}

The research for this work was supported in part by the National Research Foundation (NRF) of South Africa (Grant Number: 118532).

\appendix

\section{Wigner functional of a linear process}\label{linop}

Consider a linear operation for a single-photon state, expressed as a generalization of the single-photon projection operator, given by
\begin{equation}
\hat{P}^{(1)} = \int\ket{\mathbf{k}} P(\mathbf{k},\mathbf{k}') \bra{\mathbf{k}'}\ {\rm d}k\ {\rm d}k' ,
\end{equation}
where $P(\mathbf{k},\mathbf{k}')$ represents the linear transformation to be performed on the angular spectrum (Fourier domain wave function) of the photon and $\ket{\mathbf{k}}$ denotes the elements of a momentum basis. The generalization to arbitrary numbers of particles can then be represented as a sum of tensor products
\begin{eqnarray}
\hat{P} & = \sum_{m=0}^{\infty} \frac{1}{m!} \left[\int\ket{\mathbf{k}} P(\mathbf{k},\mathbf{k}') \bra{\mathbf{k}'}\ {\rm d}k\ {\rm d}k'\right]^{\otimes m} \nonumber \\
& = \exp_{\otimes}\left[\int\ket{\mathbf{k}} P(\mathbf{k},\mathbf{k}') \bra{\mathbf{k}'}\ {\rm d}k\ {\rm d}k'\right] ,
\label{fotop}
\end{eqnarray}
where $\exp_{\otimes}(\cdot)$ is defined such that all products in its expansion are tensor products and the first term in the expansion involves only the vacuum state $\ket{\rm vac}\bra{\rm vac}$.

To represent the operator as a Wigner functional, we use a coherent state assisted approach \cite{wigfunk}. It requires that the spatial degrees of freedom be represented in the Fourier domain, as in (\ref{fotop}). Then, we need to compute the overlap of the operator with coherent states of both sides
\begin{eqnarray}
\bra{\alpha_1}\hat{P}\ket{\alpha_2} & = & \sum_{m=0}^{\infty} \frac{1}{m!} \bra{\alpha_1}\left[\int\ket{\mathbf{k}} P(\mathbf{k},\mathbf{k}') \bra{\mathbf{k}'}\ {\rm d}k\ {\rm d}k' \right]^{\otimes m}\ket{\alpha_2} \nonumber \\
 & = & \sum_{m=0}^{\infty} \frac{1}{m!}\int \bra{\alpha_1}\left[\prod_p^m a^{\dag}(\mathbf{k}_p)\right]\ket{\rm vac} \left[\prod_p^m P(\mathbf{k}_p,\mathbf{k}_p')\right] \nonumber \\
& & \times \bra{\rm vac}\left[\prod_p^m a(\mathbf{k}_p')\right]\ket{\alpha_2} \prod_p^m\ {\rm d}k_p\ {\rm d}k_p' \nonumber \\
& = & \braket{\alpha_1}{\rm vac}\braket{\rm vac}{\alpha_2} \sum_{m=0}^{\infty} \frac{1}{m!} \left(\alpha_1^*\diamond P\diamond \alpha_2\right)^m \nonumber \\
& = & \exp \left(-\frac{1}{2}||\alpha_1||^2-\frac{1}{2}||\alpha_2||^2+\alpha_1^*\diamond P\diamond \alpha_2 \right) ,
\label{apa}
\end{eqnarray}
where $||\alpha||^2\equiv\alpha^*\diamond\alpha$.

We now use the result in (\ref{apa}) to compute the Wigner functional for a general linear process. Substituting it into the expression for the Wigner functional (see \cite{wigfunk}), and performing the functional integrations over $\alpha_1$ and $\alpha_2$, we obtain
\begin{eqnarray}
W_{P}[q,p] & = & \int \exp\left(- 2||\alpha||^2+2\alpha^*\diamond\alpha_1+2\alpha_2^*\diamond\alpha-||\alpha_1||^2 -||\alpha_2||^2 \right. \nonumber \\
& & \left. -\alpha_2^*\diamond\alpha_1 + \alpha_1^*\diamond P\diamond \alpha_2 \right)\ {\cal D}[\alpha_1,\alpha_2] \nonumber \\
& = & \frac{1}{{\rm det}\left\{\mathbf{1}+P\right\}} \exp\left[- 2\alpha^*\diamond (\mathbf{1}-P)\diamond(\mathbf{1}+P)^{-1}\diamond\alpha \right] ,
\end{eqnarray}
where it is assumed that $(\mathbf{1}+P)$ is invertable and
\begin{equation}
{\rm det}\left\{\mathbf{1}+P\right\} \equiv \exp\left[{\rm tr}\left\{\ln\left(\mathbf{1}+P\right)\right\}\right] .
\end{equation}

\section{Generating function for fixed-spectrum Fock states}
\label{genfock}

Here, we use the coherent state assisted approach \cite{wigfunk} to derive a generating function for the Wigner functionals of the fixed-spectrum Fock states. The latter are defined as
\begin{equation}
\ket{n_F} = \frac{1}{\sqrt{n!}}\left(F\diamond\hat{a}^{\dag}\right)^n \ket{\rm vac} ,
\label{fsfsdef}
\end{equation}
where $F$ is a normalized complex-valued parameter function, and $\hat{a}^{\dag}$ is the creation operator. The overlap between the density operator for such a Fock state and two fixed-spectrum coherent states is
\begin{equation}
\braket{\alpha_1}{n_F} \braket{n_F}{\alpha_2} = \exp \left(-\frac{1}{2}||\alpha_1||^2-\frac{1}{2}||\alpha_2||^2\right)
\frac{1}{n!} (\langle\alpha_1,F\rangle \langle F,\alpha_2\rangle)^n ,
\end{equation}
where $\langle\alpha_1,F\rangle$ represents an inner product between the parameter functions of the fixed-spectrum coherent state and the fixed-spectrum Fock state, respectively. We represent the overlap in terms of a generating function
\begin{equation}
{\cal K} = \exp\left(-\frac{1}{2}||\alpha_1||^2-\frac{1}{2}||\alpha_2||^2+\eta \langle \alpha_1,F \rangle \langle F,\alpha_2 \rangle \right) ,
\label{ketadef}
\end{equation}
where $\eta$ is an auxiliary parameter. The overlap is recovered by
\begin{equation}
\braket{\alpha_1}{n_F} \braket{n_F}{\alpha_2} = \left. \frac{1}{n!} \partial_{\eta}^n {\cal K} \right|_{\eta=0} .
\end{equation}
Substituting $\bra{\alpha_1}\hat{A}\ket{\alpha_2}\rightarrow {\cal K}$ into the expression for the Wigner functional (see \cite{wigfunk}), we obtain a functional integral expression for the generating function of the Wigner functionals of fixed-spectrum Fock states, given by
\begin{eqnarray}
{\cal W}(\eta) = & {\cal N}_0 \int \exp\left(- 2\alpha^*\diamond\alpha+2\alpha^*\diamond\alpha_1+2\alpha_2^*\diamond\alpha-\alpha_1^*\diamond\alpha_1 \right. \nonumber \\
 & \left. -\alpha_2^*\diamond\alpha_2-\alpha_2^*\diamond\alpha_1
 +\eta\alpha_1^*\diamond F F^*\diamond\alpha_2\right)\ \Dcirc[\alpha_1,\alpha_2] .
\label{fockwig}
\end{eqnarray}
After performing the functional integrations over $\alpha_1$ and $\alpha_2$, we get
\begin{equation}
\fl {\cal W}(\eta) = \frac{{\cal N}_0}{\det\{\mathbf{1}+\eta F F^*\}} \exp\left( -2||\alpha||^2+4\eta\alpha^*\diamond F F^*\diamond \left(\mathbf{1}+\eta F F^*\right)^{-1}\diamond\alpha\right) .
\end{equation}
Thanks to the fact that the parameter functions of the fixed-spectrum Fock states are normalized ($F\diamond F^* = 1$), one can show that
\begin{equation}
\det\{1+\eta F F^*\} = 1+\eta ,
\end{equation}
and
\begin{equation}
\left(\mathbf{1}+\eta F F^*\right)^{-1} = \mathbf{1}+\frac{\eta}{1+\eta} F F^* .
\end{equation}
It then follows that
\begin{equation}
{\cal W}(\eta) = \frac{{\cal N}_0}{1+\eta} \exp\left(-2||\alpha||^2+\frac{4\eta}{1+\eta}|\langle\alpha,F\rangle|^2\right) ,
\label{fockwig2}
\end{equation}
where ${\cal N}_0$ is the normalization constant. The Wigner functionals for the individual Fock states are given by
\begin{equation}
W_{\ket{n}\bra{n}}[q,p] = {\cal N}_0 (-1)^n L_n\left(4|\langle\alpha,F\rangle|^2\right)\exp\left(-2||\alpha||^2\right) ,
\label{fockwign}
\end{equation}
where $L_n(\cdot)$ represents the Laguerre polynomial of order $n$.

\section*{References}


\providecommand{\newblock}{}

\end{document}